*Insight*
**Lessons From the Physics-Education Reform Effort**[1]
*Richard R. Hake*

*Indiana University*

- Abstract
- Introduction
- Survey Summary
- Fourteen Lessons

## ABSTRACT

Several years ago I reported a survey (Hake 1998a,b,c) of pre/post test data for 62 introductory physics courses enrolling a total of 6542 students. The present article provides a summary of that survey and presents fourteen lessons from the physics-education reform effort that may assist the general upgrading of education and science literacy.

## I. INTRODUCTION

For over three decades, physics-education researchers repeatedly showed that *Traditional* (T) introductory physics courses with passive-student lectures, recipe labs, and algorithmic problem exams were of limited value in enhancing conceptual understanding of the subject (McDermott & Redish 1999). Unfortunately, this work was largely ignored by the physics and education communities until Halloun & Hestenes (HH) (1985a,b) devised the *Mechanics Diagnostic* (MD) test of conceptual understanding of Newtonian mechanics. Among the virtues of the MD, and the subsequent *Force Concept Inventory* (FCI) (Hestenes et al. 1992, Halloun et al. 1995) tests, are: (a) the multiple-choice format facilitates relatively easy administration of the tests to thousands of students, (b) the questions probe for conceptual understanding of basic concepts of Newtonian mechanics in a way that is understandable to the novice who has never taken a physics course (and thus can be given as an introductory-course pre-test), while at the same time rigorous enough for the initiate.

---

[1] Submitted on 19 August 2001 to *Conservation Ecology* < http://www.consecol.org/Journal/ >, a "peer-reviewed journal of integrative science and fundamental policy research."

A *typical* HH-type question is as follows (an actual HH question is avoided to help preserve the confidentiality of the test):

> A student in a lab holds a brick of weight W in her outstretched horizontal palm and lifts the brick vertically upward at a constant speed. While the brick is moving vertically upward at a constant speed, the magnitude of the force on the brick by the student's hand is:
> A. constant in time and zero.
> B. constant in time, greater than zero, but less than W.
> C. constant in time and W.
> D. constant in time and greater than W.
> E. decreasing in time but always greater than W.

Note that the responses include as distractors not only "D," the common Aristotelian misconception that "motion requires a net force," but also other less common student misconceptions "A" and "E" that might not be known to traditional teachers. Unfortunately, too few teachers "shut up and listen to their students" so as to find out what they are thinking (Arons 1981). The distractors are based on my years of listening to students as they worked through the experiments in *Socratic Dialogue Inducing Lab* #1 "Newton's First and Third Laws" (Hake 2001a). For *actual* HH questions the distractors were usually gleaned through careful *qualitative* research involving interviews with students and the analysis of their oral and written responses to mechanics questions.

Using the *Mechanics Diagnostic* test, Halloun & Hestenes (1985a,b) published a careful study using massive pre- and post-course testing of students in both calculus and non-calculus-based introductory physics courses at Arizona State University, and concluded that:
> (1) ". . . . the student's initial qualitative, common-sense beliefs about motion and . . . .(its) . . . . causes have a large effect on performance in physics, but conventional instruction induces only a small change in those beliefs.
> (2) Considering the wide differences in the teaching styles of the four professors . . . . (involved in the study) . . . . the basic knowledge gain under conventional instruction is essentially independent of the professor."

These outcomes were consistent with work done prior to the HH study as recently reviewed by McDermott & Redish (1999).

The HH results stimulated a flurry of research and development aimed at improving introductory mechanics courses. Most of the courses so generated sought to promote conceptual understanding through use of pedagogical methods published by physics-education researchers (see, e.g., Physical Science Resource Center 2001, Galileo Project 2001, UMd-PERG 2001a). These methods are usually based on the insights of cognitive science (Gardner 1985; Mestre & Touger 1989; Redish 1994; Bruer 1994, 1997; Bransford et al. 1999; Donovan et al. 1999) and/or outstanding classroom teachers (e.g., Karplus 1977, 2001; Minstrell 1989; Arons 1990; McDermott 1991, 1993; Fuller 1993; Reif 1995; Wells et al. 1995; Zollman 1996; Laws 1997). Although the methods differ in detail, they all attempt to guide students to *construct* their understandings by heads-on (always) and hands-on (usually) activities that yield immediate feedback through discussion with peers and/or instructors [*Interactive Engagement* (IE)], so as to finally arrive at the viewpoint of the professional physicist.

The survey summarized below documents some of the successes and failures of courses employing IE methods, may assist a much needed further improvement in introductory mechanics instruction in the light of practical experience, and may serve as a model for promoting educational reform in other disciplines. [Since the summary omits some important aspects, serious education scholars are urged to consult the original sources (Hake 1998a,b,c).] I then present fourteen somewhat subjective lessons from my own interpretation of the physics-education reform effort with the hope that they may assist the general upgrading of education and science literacy.

## II. SURVEY SUMMARY

Starting in 1992, I requested that pre/post-FCI data and post-test Mechanics Baseline (a problem-solving test due to Hestenes & Wells, 1992) data be sent to me. Since instructors are more likely to report higher-gain courses, the detector is biased in favor of those courses, but can still answer a crucial research question: *Can the use of Interactive Engagement (IE) methods increase the effectiveness of introductory mechanics courses well beyond that obtained by traditional methods?*

**A. The Data**

Figure 1 shows data from the survey (Hake 1998a,b,c) of 62 introductory physics courses enrolling a total 6542 students. The data are derived from pre/post scores of the MD and FCI tests indicated above, recognized for high validity and consistent reliability (for the technical meaning of these terms see, e.g., Light et al. 1990, Slavin 1992, Beichner 1994). Average pre/post test scores, standard deviations, instructional methods, materials used, institutions, and instructors for each of the survey courses are tabulated and referenced in Hake (1998b). The latter paper also gives case histories for the seven IE courses whose effectiveness as gauged by pre-to-post test gains was close to those of T courses, advice for implementing IE methods, and suggestions for further research. Various criticisms of the survey (and physics-education research generally) are countered by Hake (1998c).

For survey classification and analysis purposes I *operationally* defined:

   1. *Interactive Engagement* (IE) <u>methods</u> as those designed at least in part to promote conceptual understanding through interactive engagement of students in heads-on (always) and hands-on (usually) activities which yield immediate feedback through discussion with peers and/or instructors, all as judged by their literature descriptions;

   2. IE <u>courses</u> as those reported by instructors to make substantial use of IE methods;

   3. *Traditional* (T) <u>courses</u> as those reported by instructors to make little or no use of IE methods, relying primarily on passive-student lectures, recipe labs, and algorithmic-problem exams.

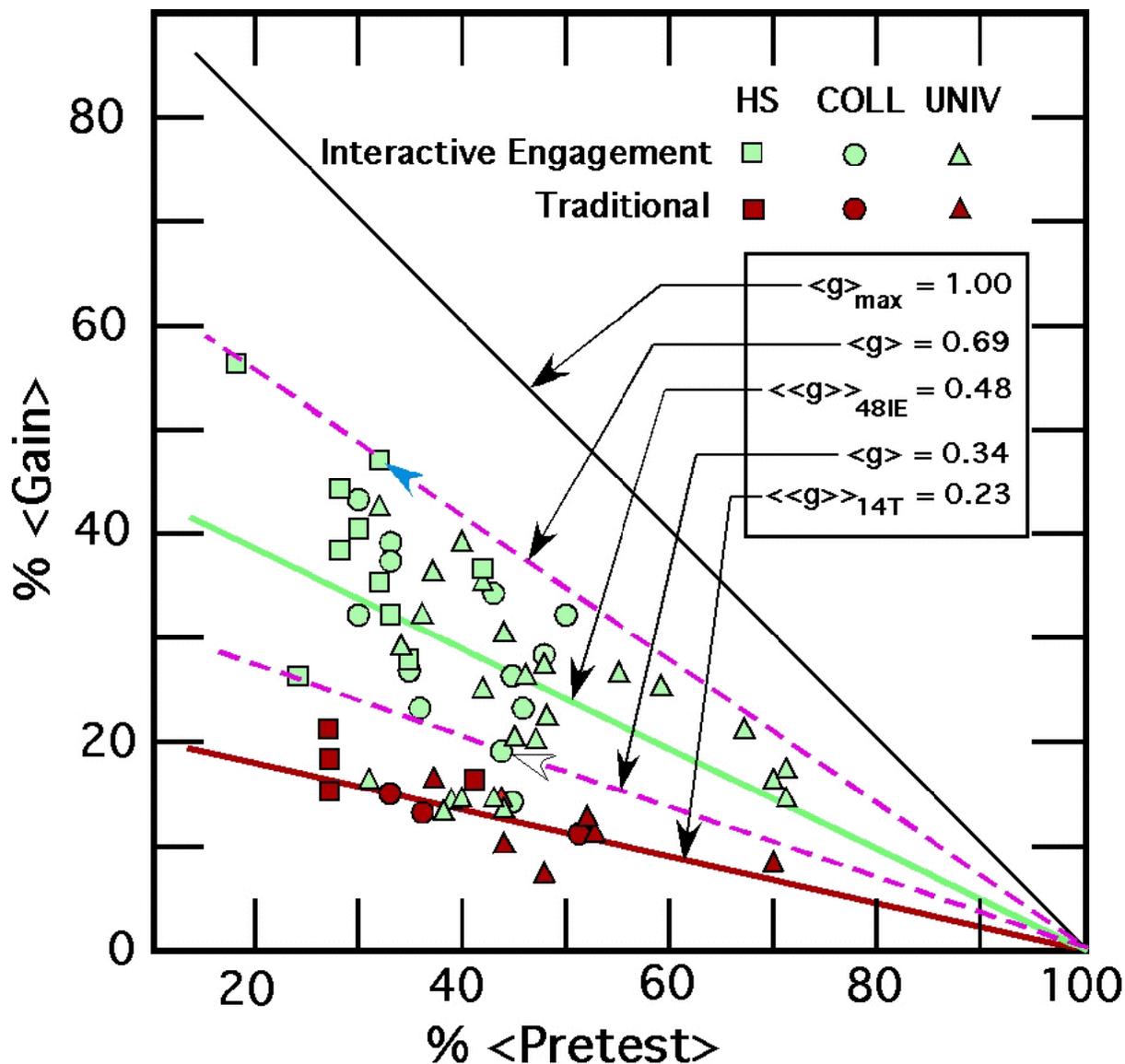

Fig. 1. The %<Gain> vs %<Pretest> score for 62 courses enrolling a total of 6542 students. Here %<Gain> = %<posttest> – %<pretest>, where the angle brackets "<....>" indicate an *average* over all students in the course. Points for high school (HS), college (COLL), and university (UNIV) courses are shown in green for *Interactive Engagement* (IE), and in red for *Traditional* (T) courses. The straight negative-slope lines are lines of constant *average normalized gain* <g>. The two dashed purple lines show that most IE courses achieved <g>'s between 0.34 and 0.69. The definition of <g>, and its justification as an index of course effectiveness, is discussed in the text. The average of <g>'s for the 48 IE courses is <<g>>$_{48IE}$ = 0.48 ± 0.14 (standard deviation) while the average of <g>'s for the 14 T courses is <<g>>$_{14T}$ = 0.23 ± 0.04 (sd). Here the double angle brackets "<<....>>" indicate an *average of averages*. (Same data points and scales as in Fig. 1 of Hake 1998a.)

A histogram of the data of Fig. 1 is shown in Fig. 2.

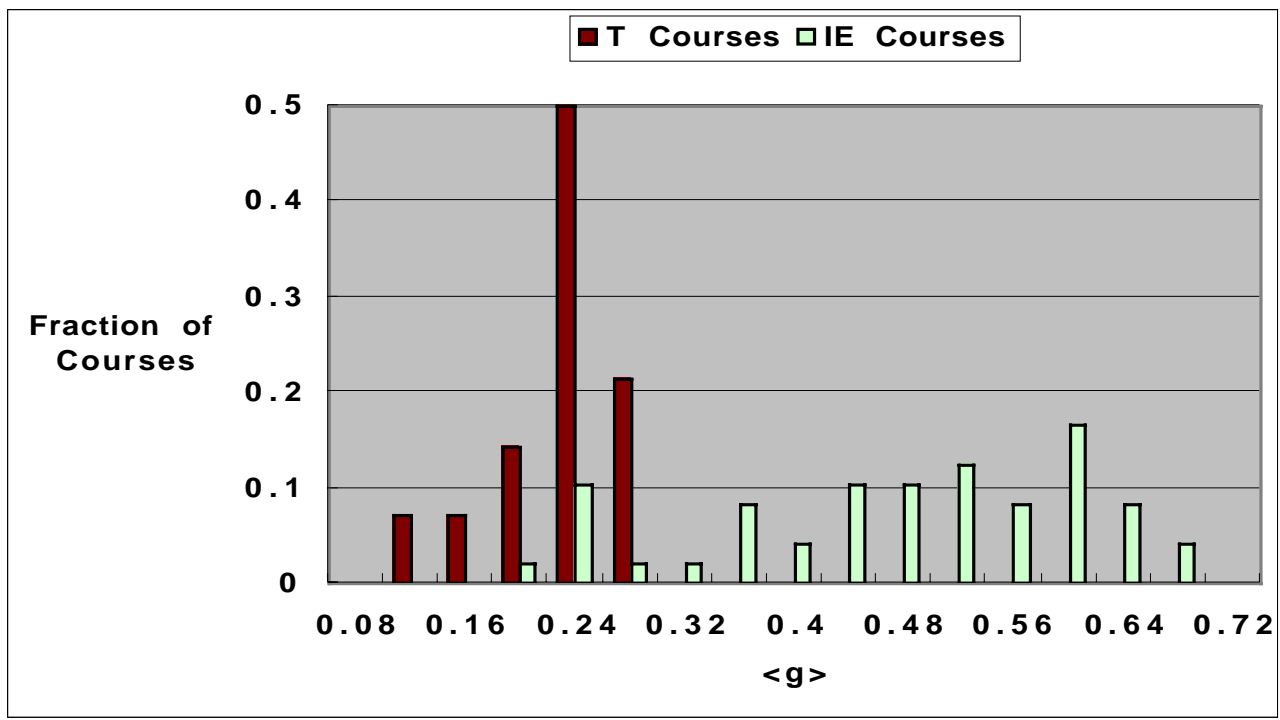

Fig. 2. Histogram of the average normalized gain <g>: red bars show the *fraction* of 14 Traditional (T) courses (2108 students) and green bars show the *fraction* of 48 Interactive Engagement (IE) courses (4458 students), both within bins of width <g> = 0.04, centered on the <g> values shown. (Same as Fig. 2 of Hake 1998a.)

## B. Average Normalized Gain

In the survey (Hake 1998a,b,c) it was useful to discuss the data in terms of a quantity which I called the "average *normalized* gain" $<g>$, *defined* as the actual gain, %$<Gain>$, divided by the maximum possible actual gain, %$<Gain>_{max}$:

$$<g> \ = \ \%<Gain> \ / \ \%<Gain>_{max} \quad \ldots \ldots \ldots \ (1a)$$

$$<g> \ = \ (\%<posttest> - \%<pretest>) \ / \ (100 - \%<pretest>) \ \ldots \ldots \ldots \ (1b)$$

where %$<posttest>$ and %$<pretest>$ are the final (posttest) and initial (pretest) class percentage averages.

For example, suppose that for a given class the pre-instruction test average was %$<pretest>$ = 44%, and the post-instruction test average was %$<posttest>$ = 63%. Then the percentage average actual gain

$$\%<Gain> = 63\% - 44\% = 19\%.$$

The *maximum* possible actual gain for this class would have been

$$\%<Gain>_{max} = (100\% - 44\%) = 56\%.$$

Thus, for this example, the average *normalized* gain

$$<g> \ = \ \%<Gain> \ / \ \%<Gain>_{max} = \ 19\%/56\% = 0.34,$$

that is, the class made an average gain of 34% of the maximum possible average gain.

To understand the graphical interpretation of the "average *normalized* gain" $<g>$, consider the same example as above. Data for that class would be plotted in Fig. 1 as the point [%$<pretest>$ = 44%, %$<Gain>$ = 19%] at the tip of the <u>white arrowhead.</u> This point has an abscissa (100% - 44%) = 56% and ordinate 19%. The absolute value of the slope "s" of the <span style="color:purple">purple dashed line</span> connecting this point to the lower right vertex of the graph is |s| = ordinate/abscissa = 19%/56% = 0.34. Thus, this absolute slope
|s| = %$<Gain>$/(100% – %$<pretest>$) = %$<Gain>$/ (maximum possible %$<Gain>$) = %$<Gain>$/ %$<Gain>_{max}$ is, of course, just the "average *normalized* gain" $<g>$. That $<g>$ can be taken to be *an index of that course's effectiveness* is justified below in Sec. II-D1. Thus, all courses with points close to the lower <span style="color:purple">purple dashed line</span> are judged to be of about equal average effectiveness, regardless of their average pretest scores. A similar calculation for the point [%$<pretest>$ = 32%, %$<Gain>$ = 47%] at the tip of the *<span style="color:blue">blue</span>* arrowhead, yields $<g>$ = 0.69. The *maximum* value of $<g>$ occurs when the %$<Gain>$ is equal to %$<Gain>_{max}$ and is therefore 1.00, as shown in Fig. 1.

**C. Popular Interactive Engagement Methods**

For the 48 interactive-engagement courses of Figs. 1 & 2, the ranking in terms of number of IE courses using each of the more popular methods is as follows:

(1) *Collaborative Peer Instruction* (Johnson et al. 1991; Heller et al. 1992a,b; Slavin 1995; Johnson et al. 2000): 48 (all courses) {CA} – for the meaning of "CA," and similar abbreviations below within the curly brackets "{…}", see the paragraph following this list.

(2) *Microcomputer-Based Labs* (Thornton and Sokoloff 1990, 1998): 35 courses {DT}.

(3) *Concept Tests* (Mazur 1997, Crouch & Mazur 2001): 20 courses {DT}; such tests for physics, biology, and chemistry are available on the web along with a description of the *Peer Instruction* method at the Galileo Project (2001).

(4) *Modeling* (Halloun & Hestenes 1987; Hestenes 1987, 1992; Wells et al. 1995): 19 courses {DT + CA}; a description is on the web at < http://modeling.la.asu.edu/ >.

(5) *Active Learning Problem Sets* or *Overview Case Studies* (Van Heuvelen 1991a,b; 1995): 17 courses {CA}; information on these materials is online at < http://www.physics.ohio-state.edu/~physedu/ >.

(6) Physics-education-research based text (referenced in Hake 1998b, Table II) or no text: 13 courses.

(7) *Socratic Dialogue Inducing Labs* (Hake 1987, 1991, 1992, 2001a; Tobias & Hake 1988): 9 courses {DT + CA}; a description and lab manuals are on the web at the Galileo Project (2001) and < http://www.physics.indiana.edu/~sdi >.

The notations within the curly brackets {. . .} follow Heller (1999) in loosely associating the methods with "learning theories" from cognitive science. Here "DT" stands for "Developmental Theory," originating with Piaget (Inhelder & Piaget 1958, Gardner 1985, Inhelder et al. 1987, Phillips & Soltis 1998); and "CA" stands for "Cognitive Apprenticeship" (Collins et al. 1989, Brown et al. 1989). All the methods (save #6) recognize the important role of social interactions in learning (Vygotsky 1978, Lave & Wenger 1991, Dewey 1997, Phillips & Soltis 1998). It should be emphasized that the above rankings are by popularity within the survey, and have no necessary connection with the effectiveness of the methods relative to one another. In fact, it is quite possible that some of the less popular methods used in some survey courses, as listed by Hake (1998b), could be more effective in terms of promoting student understanding than any of the above popular strategies.

**D. Conclusions of the Survey**

The conclusions of the survey (Hake 1998a,b,c; 1999a) may be summarized as follows:

1. **The average normalized gain <g> affords a consistent analysis of pre/post test data on conceptual understanding over diverse student populations in high schools, colleges, and universities**. For the 62 courses of the survey (Hake 1998a,b,c) the correlation of

$$<g> \text{ with } (\%<\text{pretest}>) \text{ is } + 0.02. \quad \dots \dots \dots \dots \dots \dots \dots \dots \quad (2)$$

This constitutes an experimental justification for the use of <g> as a comparative measure of course effectiveness over diverse student populations with widely varying average pretest scores, and is a reflection of the usually relatively small correlations of single student g's with their pretest scores within a given class (Hake 1998a, 2001b; Cummings 1999; Meltzer 2001).

The average posttest score (%<postest>) and the average actual gain (%<Gain>) are less suitable for comparing course effectiveness over diverse groups since their correlations with (%<pretest) are significant. The correlation of

$$(\%<\text{posttest}>) \text{ with } (\%<\text{pretest}>) \text{ is } + 0.55, \quad \dots \dots \dots \dots \dots \dots \dots \dots \quad (3)$$

and the correlation of

$$(\%<\text{Gain}>) \text{ with } \%<\text{pretest}> \text{ is } – 0.49, \quad \dots \dots \dots \dots \dots \dots \dots \dots \quad (4)$$

both of which correlations would be anticipated. Note that in the *absence* of instruction, a high positive correlation of (%<posttest >) with (%<pretest>) would be expected. The successful use of the normalized gain for the analysis of pre/post test data in this and other physics-education research (see Sec. II-G), calls into question the common dour appraisals of pre/post test designs (Lord 1956, 1958; Cronbach & Furby 1970; Cook & Campbell 1979). For a review of the pre/post literature (pro and con) see Wittmann (1997).

2. Fourteen *Traditional* (T) courses (2084 students) of the survey yielded

$$<<g>>_{14T} = 0.23 \pm 0.04 \text{sd} \quad \dots \dots \dots \dots \dots \dots \dots \dots \dots \dots \dots \dots \quad (5)$$

Considering the *elemental* nature of the MD/FCI questions (many physics teachers regard them as too easy to be used on examinations) and the relatively low <g> = 0.23 (i.e., *only 23% of the possible gain was achieved*), it appears that **traditional (T) courses fail to convey much basic conceptual understanding of Newtonian mechanics to the average student**.

3. Forty-eight *Interactive Engagement* (IE) courses (4458 students) of the survey yielded

$$\langle\langle g\rangle\rangle_{48IE} = 0.48 \pm 0.14\text{sd} \dots\dots\dots\dots\dots\dots\dots\dots\dots\dots\dots\dots\dots(6)$$

The $\langle\langle g\rangle\rangle_{48IE}$ is over twice that of $\langle\langle g\rangle\rangle_{14T}$. The difference ($\langle\langle g\rangle\rangle_{48IE} - \langle\langle g\rangle\rangle_{14T}$) is over 6 sd's of $\langle\langle g\rangle\rangle_{14T}$ and almost two sd's of $\langle\langle g\rangle\rangle_{48IE}$, reminiscent of differences seen in comparing instruction delivered to students in large groups with one-on-one instruction (Bloom 1984). **This suggests that IE courses can be much more effective than T courses in enhancing conceptual understanding of Newtonian mechanics.** Although it was not possible in this survey to randomly assign students from a single large homogeneous population to the T and IE courses, contrasting T and IE data are all drawn from the same institutions and the same generic introductory program regimes (Hake 1998b). Thus it seems very unlikely that the nearly-two-sd difference between $\langle\langle g\rangle\rangle$'s for the IE and T courses could be accounted for by differences in the student populations.

An alert critic of an early draft [and more recently Becker (2001a), Sec. II-E3 below] have pointed out that the $\langle\langle g\rangle\rangle$ difference might be due in part to the apparent smaller average enrollment for IE courses (4458/48 = 93) than for T courses (2084/14 = 149). However such calculation of average *class* size is invalid because in several cases (Hake 1998b, Table I) classes of fairly homogeneous instruction and student population were combined into one "course" whose $\langle g\rangle$ was calculated as a number-of-student-weighted average of the $\langle g\rangle$'s of the individual classes. A correct calculation yields an average *class* enrollment of 4458/63 = 71 for IE classes and 2084/34 = 61 for T classes, so the average class sizes are quite similar.

4. A detailed analysis of random and systematic errors has been carried out (Hake 1998a) but will not be repeated here. Possible systematic errors considered were: (a) question ambiguities and isolated false positives (right answers for the wrong reasons); and uncontrolled variables in the testing conditions such as (b) teaching to the test and test-question leakage, (c) the fraction of course time spent on mechanics, (d) post and pretest motivation of students, and (f) the Hawthorne/John Henry effects. It was concluded that "**it is extremely unlikely that random or systematic error plays a significant role in the nearly two-standard deviation difference in the $\langle\langle g\rangle\rangle$'s of T and IE courses.**"

5. Conclusions 1–3 above are bolstered by an analysis (Hake 1999a) of the survey data in terms of Cohen's (1988) "effect size" d. The effect size is commonly used in meta-analyses (e.g., Light et al. 1990, Hunt 1997, Glass 2000), and strongly recommended by many psychologists (B. Thompson 1996, 1998, 2000), and biologists (Johnson 1999, Anderson et al. 2000, W.L. Thompson 2001) as a preferred alternative (or at least addition) to the usually inappropriate (Rozeboom 1960, Carver 1993, Cohen 1994, Kirk 1996) t-tests and p values associated with null-hypothesis testing.

Carver (1993) subjected the Michelson & Morley (1887) data to a simple analysis of variance (ANOVA) and *found **statistical** significance associated with the direction the light was traveling (p < 0.001*)! He writes: "It is interesting to speculate how the course of history might have changed if Michelson and Morley had been trained to use this *corrupt form of the scientific method*, that is, testing the null hypothesis first. They might have concluded that there was evidence of *significant* differences in the speed of light associated with its direction and that therefore there was evidence for the luminiferous ether . . . . Fortunately Michelson and Morley . . .(first). . . .interpreted their data with respect to their research hypothesis." (My *italics*.) Consistent with the scientific methodology of physical scientists such as Michelson/ Morley (see Sec. II-G), Rozeboom (1960) wrote: "*. . . the primary aim of a scientific experiment is not to precipitate decisions, but to make an appropriate adjustment in the degree to which one accepts, or believes, the hypothesis or hypotheses being tested.*" (See also Anderson 1998.)

The effect size d is defined by Cohen (1988, p. 20, 44) as

$$d = |m_A - m_B| / [(sd_A^2 + sd_B^2)/2]^{0.5} \dots \dots \dots \dots \dots \dots \dots \dots \dots \dots (7)$$

where $m_A$ and $m_B$ are population means expressed in the raw (original measurement) unit, and where the denominator is the root mean square of standard deviations for the A- and B-group means, sometimes called the "pooled standard deviation." For the present survey Eq. (7) becomes

$$d = [<<g>>_{48IE} - <<g>>_{14T}] / [(sd_{48IE}^2 + sd_{14T}^2)/2]^{0.5} \dots \dots \dots \dots \dots \dots (8)$$

Insertion of the measured values $<<g>>_{14T} = 0.23 \pm 0.04\,sd$ and $<<g>>_{48IE} = 0.48 \pm 0.14\,sd$ into Eq. (8) yields:

$$d = [(0.48) - (0.23)] / [(0.14^2 + 0.04^2)/2]^{0.5} = 2.43 \dots \dots \dots (9)$$

The above "d" can be compared with:

(a) Cohen's (1988, p. 24) rule of thumb – based on typical results in social science research – that d = 0.2, 0.5, 0.8 imply respectively "small," "medium," and "large" effects. But Cohen cautions that the adjectives "are relative, not only to each other, but to the area of behavioral science or even more particularly to the specific content and research method being employed in any given investigation."

(b) The course-enrollment N-weighted <d> = 0.57 obtained for 31 test/control group studies (2559 students) of achievement by Springer et al. (1999, Table 2) in a meta-analyisis of the effect of small-group learning among undergraduates in science, math, and engineering.

The effect size "d" of the present study is much larger than might be expected on the basis of Cohen's rule of thumb, or on the basis of the results of Springer et al. This difference may be related to the facts that in this survey, unlike most education-research meta-analyses (e.g., that of Springer et al., Slavin 1995, and Johnson et al. 2000): (1) *all* courses covered nearly the same material (here introductory Newtonian mechanics); (2) the material is conceptually difficult and counterintuitive; (3) the *same* test (either MD or FCI – see Sec. I ) was administered to both IE and T classes; (4) the tests employed are widely recognized for their validity and consistent reliability, have been carefully designed to measure understanding of the key concepts of the material, and are far superior to the plug-in-regurgitation type tests so commonly used as measures of "achievement"; (5) the measurement unit gauges the normalized learning *gain* from start to finish of a course, *not* the "achievement" at the end of a course; (6) the measurement unit <g> is not significantly correlated with students initial knowledge of the material being tested; (7) the "treatments" are all patterned after those *published by education researchers in the discipline being tested*. I think that the Springer et al. meta-analysis probably understates the potential of small-group learning for advancing conceptual understanding and problem-solving ability.

For what it's worth, the conventional t-test (Slavin 1992, p. 157-162; Snedecor & Cochran 1989, p. 83-102) [which assumes *normal* distributions, unlike those presently observed (Fig. 2)], yields t = 11 and p < 0.001. Paraphrasing Thompson (1996), "p" is the two-tailed probability (0 to 1.0) of the results (means difference and the sd's), given the sample sizes and *assuming the samples were derived from a population in which the null hypothesis $H_0$ is exactly true*, i.e., the probability of the results assuming that $H_0$ is true. For discussions of the common misinterpretation of "p" as the converse (the probability that $H_0$ is true given the results), see Cohen (1994) and Kirk (1996). [Because the variances of the <g> distributions for the IE and T courses are markedly dissimilar (F = $sd_{IE}^2/sd_T^2$ = 12), I have employed an approximation to the standard t-test due to Satterthwaite (1946), as discussed by Snedecor & Cochran (1989, p. 96-98). and by Rosenthal et al. (2000, p. 33-35).]

An effect size d can also be calculated directly for each pre/post study in accord with Eq. (10):

$$d = (\%\langle post \rangle - \%\langle pre \rangle) / [(sd_{pre}^2 + sd_{post}^2)/2]^{0.5} \quad \ldots \ldots \ldots \ldots \ldots \ldots \ldots (10)$$

However, it should be noted that for pre/post comparisons of course effectiveness over diverse student populations with widely varying average pretest scores, "d" is a relatively poor metric because unlike "$\langle g \rangle$" : (a) "d" depends on the actual bare (unrenormalized) actual average %$\langle$Gain$\rangle$, whose magnitude, as indicated in Eq. (4), tends to be negatively correlated with the %$\langle$pretest$\rangle$; (b) %$\langle$Gain$\rangle$'s are confounded with sd's: given two classes both with identical *statistically* significant %$\langle$Gain$\rangle$'s, the more homogeneous class with the smaller ($sd_{pre}^2 + sd_{post}^2$) will be awarded the higher "d." Ignoring these problems, for the 33 courses of the survey for which sd's are available (Hake 1998b), I obtain average effect sizes:

$\langle d \rangle_{9T} = 0.88 \pm 0.32$sd for 9T courses (1620 students) with $\langle\langle g \rangle\rangle_{9T} = 0.24 \pm 0.03$sd; and

$\langle d \rangle_{24IE} = 2.16 \pm 0.75$sd for 24 IE courses (1843 students) with $\langle\langle g \rangle\rangle_{24IE} = 0.50 \pm 0.12$sd.

Weighting the d and g averages, as in the Springer et al. (1999) study, in accord with class enrollments N (on the grounds that d and $\langle g \rangle$ from larger classes more closely approximate actual effects in very large student populations) yields:

$\langle d \rangle_{9T\text{-Nweighted}} = 1.05$ for 9T courses (1620 students) with $\langle\langle g \rangle\rangle_{9T\text{-Nweighted}} = 0.26$, and

$\langle d \rangle_{24IE\text{-Nwighted}} = 2.16$ for 24 IE courses (1843 students) with $\langle\langle g \rangle\rangle_{24IE\text{-Nweighted}} = 0.52$, not greatly different from the non-weighted averages.

The present $\langle d \rangle_{24IE} = 2.16$ can be compared with: (1) the similar $d_{1IE} = 1.91$ (with $\langle g \rangle_{1IE} = 0.48$) reported by Zeilik, Schau, Mattern (1998) for a single IE introductory astronomy course (331 students) given in Spring 1995 at the University of New Mexico, (2) the much smaller course-enrollment N-weighted average $\langle d \rangle = 0.30$ obtained for 6 pre/post studies (764 students) of achievement by Springer et al. (1999, Table 2).

As in Sec. II-D3 above, the critic of an early draft pointed out that $\langle d \rangle$ difference might be due in part to the apparent smaller average enrollment for IE courses (1843/24 = 77) than for T courses (1620/9 = 169). However such calculation of average *class* size is invalid for the reason given previously. A correct calculation of average class size indicates an average class enrollment of 1843/39 = 47 for IE classes and 1620/31 = 52 for T classes, so the average class enrollments are quite similar.

6. Considering the elemental nature of the MD/FCI tests, **current IE methods and their implementation need to be improved**, **since none of the IE courses achieves <g> greater than 0.69.** In fact, as can be seen in Figs. 1 & 2, seven of the IE courses (717 students) achieved <g>'s close to those of the T courses. Case histories of the seven low–<g> courses (Hake 1998b) suggest that implementation problems occurred that might be mitigated by:
- (1) apprenticeship education of instructors new to IE methods,
- (2) emphasis on the nature of science and learning throughout the course,
- (3) careful attention to motivational factors and the provision of grade incentives for taking IE-activities seriously,
- (4) recognition of and positive intervention for potential low-gain students,
- (5) administration of exams in which a substantial number of the questions probe the degree of conceptual understanding induced by the IE methods,
- (6) use of IE methods in all components of a course and tight integration of those components.

Personal experience with the Indiana IE courses and communications with most of the IE instructors in the survey suggest that similar implementation difficulties probably occurred to a greater or lesser extent in all the IE courses and are probably partially responsible for the wide spread in the <g>'s, apparent for IE courses in Figs. 1 & 2.

7. I have plotted (Hake 1998a) average post-course scores on the problem-solving Mechanics Baseline test (Hestenes & Wells, 1992) [available for 30 (3259 students) of the 62 courses of the survey] vs those on the conceptual FCI. There is a very strong positive correlation r = + 0.91 of the MB and FCI scores. This correlation and the comparison of IE and T courses at the same institution (Hake 1998a) imply that *IE methods enhance problem-solving ability*.

**E. Criticisms of the Survey**

Early criticisms of the survey have been countered in Hake 1998c. William Becker (2001a), authority on economics contributions to the assessment of educational research (Becker 1995), and executive editor of the *Journal of Economic Education* < http://www.indiana.edu/~econweb >
  in his generally censorious review of recent educational research "through the lens of theoretical statistics," raised other objections to Hake (1998a):

> 1. "The amount of variability around a mean test score for a class of 20 students versus a mean of 200 students cannot be expected to be the same. Estimation of a standard error for sample of 62. . . (courses) . . . , where each of the 62 receives an equal weight ignores this heterogeniety."

> But only the standard deviations (sd's) of the $<<g>>$'s for the 48 IE and 14T courses were given (*not* the standard errors). For example, I give $<<g>>_{48IE} = 0.48 \pm 0.14$ (sd). The spread (sd) in $<g>$ values for the IE courses is large, 29% of $<<g>>_{48IE}$. In the error analysis (Hake 1998a), I adduce evidence that the large spread in the $<g>_{IE}$ distribution is due to random errors *plus other factors*: e.g., course-to-course variations in the systematic errors and in the effectiveness of the pedagogy and/or implementation. In my opinion, attempts to take into account the heterogeniety due to course enrollment would add little to the analysis.

> The crucial point, seemingly ignored by Becker, is that the difference $(<<g>>_{48IE} - <<g>>_{14T})$ is large in comparison to the spread in the data; more technically, the "effect size," $d = 2.43$ [see Sec. II-D5, Eq. (9)] is relatively large. Hence, it is extremely unlikely that details of the $<g>$ averaging procedure would affect the conclusion that IE courses can be much more effective than T courses in enhancing conceptual understanding of Newtonian mechanics. This view is borne out by the fact that the course-enrollment N-weighted $<<g>>_{48IE\text{-Nweighted}} = 0.49$ and $<<g>>_{14T\text{-Nweighted}} = 0.24$ are very close to the non-N weighted averages $<<g>>_{48IE} = 0.48$, $<<g>>_{14T} = 0.23$.

2. "Unfortunately, the gap closing outcome measure "g" is algebraically related to the starting position of the student as reflected in the pretest: g falls as the *pretest score* rises, for *maximum score ≥ posttest score ≥ pretest score*.  Any attempt to regress a posttest minus pretest change score, or its standardized gap closing measure g, on a pretest score yields a biased estimate of the pretest effect. Hake (1998) makes no reference to this bias when he discusses his regressions and correlation of average normalized gain, average gain score and posttest score on the average pretest score. These regressions suffer from the classic errors in variables problem and regression to the mean problem associated with the use of the pretest as an explanatory index variable for unknown ability.  Even those schooled in statistics continue to overlook this regression fallacy, as called to economists' attention . . . (by) . . . Nobel laureate Milton Friedman (1992)."

The first sentence is based on Becker's partial differentiation $\partial g/\partial y = (x - Q)/(Q - y)^2$, in agreement with Hake, 1998a, footnote 45.  Here Q is the number of questions on the exam, y and x are the number of correct responses on the pretest and posttest, respectively, and $g = (x - y)/(Q - y)$. But Becker's differentiation, while mathematically correct, has little physical significance because, for a single student, x is *not,* in general, independent of y; and for a single class the average <x> is *not* independent of the average <y>.  In fact, as indicated above, for the 62 courses of the survey, correlation of the average posttest score <x> with the average pretest score <y> is +0.55, while the correlation of average normalized gain <g> with <y> is a very low +0.02.

Regarding the rest of Becker's criticism #2 above: (a) Although Becker and Chizmar & Ostrosky (1999) take the pretest score as a measure of students' "ability/aptitudes" in the course, the MD/FCI pretest scores reflect the students' "initial knowledge states," which, in my experience, have little if any connection with their "ability/aptitudes."  (b) In my opinion, "classic errors in variables problem and regression to the mean problem" do not invalidate my calculation of correlation coefficients in Eqs. (2-4), and are not responsible for a significant fraction of the nearly two-standard-deviation difference in the average normalized gains of IE and T courses found in Hake (1998a,b,c).

[As an aside, Becker relates g to the "Tobit model" (named after economics Nobel laureate James Tobin) and implies that economist Frank Ghery (1972) was the first to propose use of g, evidently unaware (as was I) that g was earlier used by psychologists Hovland et al. (1949).]

3. "When studies ignore the class size . . . (see my counters to this in Secs. II-D3 & II-E1) . . . and sample selection issues, readers should question the study's findings regardless of the sample size or diversity in explanatory variables":

   a. "Hake does not give us any indication of beginning versus ending enrollments, which is critical information if one wants to address the consequence of attrition."

Becker is probably unaware of the submitted (unpublished – physics education research lacks an archival *Physical Review* – but available on the web) companion paper (Hake 1998b - called "ref. 17a" in Hake 1998a). The data Tables Ia,b,c of Hake 1998b clearly indicate which courses were and which were not analyzed with "matched" data, i.e., data in which only posttest scores of students who had also taken the pretest were included in the average posttest score. In a majority of courses matched data *were* used. Tables Ia,b,c show no obvious dependence of <g> on whether or not the data were matched. In footnote "c" of that table, I estimate, from my experience with the pre/post testing of 1263 students at Indiana University, "that the error in the normalized gain is probably less than 5% for classes with 20 – 50 students and less that 2% for classes with more than 50 students." Saul (1998, p. 117) states ". . . . I found that the matched and unmatched results . . . from his extensive pre/post FCI studies . . . . are not significantly different." Consistent with the view that the use of some unmatched data had little influence on the survey results, an analysis of *all* the matched data of the survey (4 T courses enrolling 292 students, and 34 IE courses enrolling 3511 students) yields:

$$<<g>>_{4T\text{-matched}} = 0.24 \pm 0.05\text{sd}, \quad <<g>>_{34IE\text{-matched}} = 0.50 \pm 0.13\text{sd}; \quad \ldots \quad (11)$$

$$<<g>>_{4T\text{-matched\&Nweighted}} = 0.23, \quad <<g>>_{34IE\text{-matched\&Nweighted}} = 0.51. \quad \ldots \ldots (12)$$

The matched data $<<g>>$'s of Eqs. (11 & 12) are very close to those for the complete data set of 14 T courses enrolling 2084 students and 48 IE courses enrolling 4458 students:

$$<<g>>_{14T} = 0.23 \pm 0.04\text{sd}, \quad <<g>>_{48IE} = 0.48 \pm 0.14\text{sd} \quad \ldots \ldots (5,6)$$

b. "... if test administration is voluntary, teachers who observe that their average class score is low on the pretest may not administer the posttest. This is a problem for multi-institution studies, such as that described in Hake (1998) where instructors elected to participate, administer tests and transmit data."

Becker may have overlooked Hake's (1998a, Sec. II, "Survey Method and Objective") statement: "This mode of data solicitation . . . (voluntary submission of data by teachers). . . tends to pre-select results which are biased in favor of outstanding courses which show relatively high gains on the FCI . . . As in any scientific investigation, bias in the detector can be put to good advantage if appropriate research objectives are established. We do *not* attempt to access the *average* effectiveness of introductory mechanics courses. Instead we seek to answer a question of considerable practical interest to physics teachers: ***Can*** *the classroom use of IE methods increase the effectiveness of introductory mechanics courses well beyond that attained by traditional methods*?"

c. "... Hake (1998a) ... extol(s) the power in testing associated with large national samples . . . (but fails to). . . . fully appreciate or attempt to adjust for the many sample selection problems in generating pre- and posttest scores."

I have addressed Becker's purported "sample selection problems" above. In my opinion, Becker fails to appreciate the fact that use of the normalize gain <g> obviates the need to adjust for the pretest score. Furthermore, the observed nearly two-standard-deviation difference in the <<g>>'s of IE and T courses appears to overwhelm the smaller effects of "hidden variables" as discussed in Sec. II-F below.

4. ". . . . there is relatively strong inferential evidence . . . [evidently from Almer et al. (1998) and Chizmar & Ostrosky (1999)] . . . supporting the hypothesis that periodic use of variants of the one-minute paper (wherein an instructor stops class and asks each student to write down what he or she thought was the key point and what still needed clarification at the end of a class period) increases student learning. *Similar support could not be found for other methods.* This does not say, however, that alternative teaching techniques do not work. It simply says that *there is no compelling statistical evidence saying that they do*." (My *italics*.)

Becker omits mention of the difficulties (Snedecor & Cochran 1989, Chapter 17, "Multiple Linear Regression") in standard regression analyses (such as those of Chizmar and Ostrosky and Almer et al.) with one dependent variable Y and more than one independent variable "X": e.g., possible (1) intercorrelation of hypothesized X's, (2) non-linear relationships of X's with Y, (3) omission of important X's, (4) measurement errors in X's.

In any case, Becker evidently thinks that there is *no compelling statistical evidence* that (a) IE courses can be much more effective than T courses in enhancing conceptual understanding of Newtonian mechanics, and (b) alternative teaching techniques (save minute papers) "work." If economics faculty share Becker's belief "b", then it is little wonder that their teaching methods are "still dominated by 'chalk and talk' classroom presentations" (Becker & Watts 2000, 2001). And given that minute papers are the *only* alternative technique shown to "work," why should economics instructors accept the Becker/Watts claim (evidently regarded as unsubstantiated by Becker himself) that "their ratings. . . (will). . . go up, *along with what students learn, when students are actively involved in the classroom learning experience*" ? (My *italics*).

Becker's criteria for "compelling statistical evidence" is contained in his "11-point set of criteria that all inferential studies can be expected to address in varying degrees of detail." Becker writes: "Outside of economics education I could find no examples of education researchers checking alternative regression model specifications . . . (as). . . can be seen in Chizmar and Ostrosky (1999) . . .(and) . . . Becker and Powers (2001)." Thus it would appear that all inferential studies by non-economists fail to fully satisfy point #6 of Becker's criteria: "Multivariate analyses, which includes diverse controls for things other than exposure to the treatment that may influence outcomes [e.g., instructor differences, student aptitude . . . (called "hidden variables" in Sec. II-F) . . . .], but that cannot be dismissed by randomization (which typically is not possible in education settings)."

But if the effects of such variables are small in comparison to that induced by the treatment, and the treatment is given to many heterogeneous control and test groups as in Hake (1998a,b,c), then, in my opinion, multivariant analyses, with all their uncertainties, are of dubious value in demonstrating the efficacy of the treatment as argued in Sec. II-F below.

Furthermore, Becker, in his econocentric concentration on inferential statistics, omits criteria for "compelling evidence" that most physical scientists regard as crucial, but which are largely ignored by Economics Education Researchers (EER's): (a) careful consideration of possible *systematic* errors, (b) the presentation of *raw* unmassaged data such that experiments can be repeated and checked by other investigators, and (b) the related *extent to which the* research *conclusions are independently verified by other investigators under other circumstances so as to contribute to a community map* (see Sec. II-G).

As an aside, my own experience with minute papers (Hake 1998a, ref. 40; Hake 1998b, ref. 58 and Table IIc) is that they can constitute a significant but relatively minor segment of effective interactive engagement, consistent with the results of Chizmar and Ostrosky (1999), who find a rather miniscule minute-paper effect: an appoximately 7% increase in the posttest TUCE (Saunders 1991) score relative to the pretest TUCE score (one-tailed $p = 0.025$). (Typical of EER's, Chizmar and Ostrosky do not specify an effect size.) Becker continues the usual literature misattribution of minute papers to Wilson (1986) [and indirectly to CAT champions Angelo and Cross (1993)] rather than to Berkeley physicist Charles Schwartz (1983); see also Hake (2001c).

5. In a later communication Becker (2001b) wrote "your study does not adequately address the sample selection issues associated with the way in which subjects entered your study . . . (see my response in "3b" above). . . , the attrition from the pretest to posttest . . . (see my response in "3a" above). . . and the implications for class averages. Jim Heckman . . . .(Heckman et al. 1998, Heckman 2000, Heckman et al. 2001). . . earned a Nobel Prize in economics for his work on selection issues and it is unfortunate that you appear unaware of it and the subsequent developments in econometrics and statistics. There are many applications of Heckman's work in education research including my own recent contribution . . . (Becker & Powers 2001)."

I have reviewed Heckman (2000) and Heckman et al. (1998), and agree with Becker that Heckman's work is potentially relevant to education research. However, the study by Heckman et al. (1998) concerns the danger that members of a control group may fare differently than the members of a test group had they (the control group members) been in the test group. I fail to understand how this aspect of Heckman's work, or any other aspects cited by the Nobel committee in Heckman (2000), are relevant to the control (T) and test (IE) group participants in Hake (1998a,b,c), since the participants in the latter study were all drawn from the same generic introductory physics courses, and I have addressed above the various other selection issues with which Heckman deals.

**F. Are There Important "Hidden Variables" ?** (Peat 1997)

As indicated in Sec. II-D1, Eq. (2), for the survey of Hake (1998a,b,c) the correlation of the normalized gain $\langle g \rangle$ with %$\langle$pretest$\rangle$ is a very low +0.02. However, open research questions remain as to whether or not (a) any "hidden variables" - HV's (the averages over a class of e.g., math proficiency, spatial visualization ability, scientific reasoning skills, physics aptitude, gender, personality type, motivation, socio-economic level, ethnicity, IQ, SAT, GPA) are significantly correlated with $\langle g \rangle$, and (b) the extent to which any such correlations represent causation (Cook & Campbell 1979, Light et al. 1990, Slavin 1992).

For one course (IU94S of Hake 1998b, Table Ic, N = 166, $\langle g \rangle$ = 0.65), Hake et al. (1994) found significant average mathematics pretest-score differences between high- and low-normalized-gain students. For a later course (IU95S of Hake 1998b, Table Ic, N = 209, $\langle g \rangle$ = 0.60), Hake (1995) measured correlation coefficients between single student g's and pretest scores on mathematics and spatial visualization of, respectively, +0.32 and +0.23. More recently Meltzer (2001) reported a student-enrollment weighted correlation coefficient of +0.32 for 4 courses with total enrollment N = 219 between single student g's and math-skills pretest scores.

Preliminary work by Clement (2001) suggests a positive correlation of single student g's with a pretest (Lawson 1995) of scientific reasoning. The measurements of Halloun & Hestenes (1998) and Halloun (1997) suggest the existence of a positive correlation between single student g's and pretest scores on their *Views About Sciences Survey* (VASS). Hake (1996), Henderson et al. (1999), McCullough (2000), the Galileo Project (2001), and Meltzer (2001) have reported gender differences ($\langle g \rangle_{males}$ > $\langle g \rangle_{females}$) in $\langle g \rangle$'s for some classes. [Hake calculated a gender-difference effect size 0.58 for IU95S (see above). Meltzer calculated gender-difference effect sizes of 0.44 and 0.59 for two classes (N = 59, 78) at Iowa State University, but observed no significant gender difference in two other classes (N = 45, 37) at Southeastern Louisiana University.]

Nevertheless, the $\langle g \rangle$ dependence on the above HV's is small relative to the very strong dependence of $\langle g \rangle$ on the degree of interactive engagement (effect size 2.43, Eq. 9), and would tend to average out in multi-course comparisons over widely diverse populations in both the test and control groups. Thus, I think that it is extremely unlikely that HV effects could account for an appreciable fraction of the nearly two-standard-deviation difference in the average of $\langle g \rangle$'s for the 48 IE and 14 T courses. However, such effects could, of course, be significant in the comparison of only a few non-randomly assigned courses, as emphasized by Meltzer (2001).

**G. Can Educational Research Be *Scientific* Research?**

There has been a long standing debate over whether education research can or should be "scientific" (e.g., *pro*: Dewey 1929, 1966; Anderson et al. 1998; Bunge 1999; Redish 1999; Mayer 2000; Phillips & Burbules 2000; Phillips 2000; *con*: Lincoln & Guba 1985, Schön 1995, Eisner 1997, Lagemann 2000). In my opinion, substantive education research *must* be "scientific" in the sense indicated below. My biased prediction (Hake 2000b) is that for physics-education research, and possibly even education research generally: (a) the bloody "paradigm wars" (Gage 1989) of education research will have ceased by the year 2009, with, in Gage's words, a "productive rapprochement of the paradigms," (b) some will follow paths of pragmatism or Popper's "piecemeal social engineering" to this paradigm peace, as suggested by Gage, but (c) most will enter onto this "sunlit plain" from the path marked "scientific method" as practiced by most research scientists:

(1) "EMPIRICAL: Systematic investigation . . . (by quantitative, qualitative, or any other means) . . . of nature to find reproducible patterns in the structure of things and the ways they change (processes).

(2) THEORETICAL: Construction and analysis of models representing patterns of nature." (Hestenes 1999).

(3) "Continual interaction, exchange, evaluation, and criticism so as to build a . . . . community map." (Redish 1999).

For the presently discussed research, the latter feature is demonstrated by the fact that FCI normalized gain results for IE and T courses that are consistent with those of (Hake 1998a,b,c) have now been obtained by physics-education research (PER) groups at the Univ. of Maryland (Redish et al. 1997, Saul 1998, Redish & Steinberg 1999, Redish 1999); Univ. of Montana (Francis et al. 1998); Rennselaer and Tufts (Cummings et al. 1999); North Carolina State Univ. (Beichner et al. 1999); and Hogskolan Dalarna - Sweden (Bernhard 1999); and Carnegie Mellon Univ. (Johnson 2001). In addition, PER groups have now gone beyond the original survey in showing, for example, that (a) there may be significant differences in the effectiveness of various IE methods (Saul 1998, Redish 1999); and (b) FCI data can be analyzed so as to show the distribution of incorrect answers in a class and thus indicate common incorrect student models (Bao & Redish 2001). *Thus in physics education research, just as in traditional physics research, it is possible to perform quantitative experiments that can be reproduced (or refuted) and extended by other investigators, and thus contribute to the construction of a continually more refined and extensive "community map."*

## III. FOURTEEN LESSONS FROM THE PHYSICS-EDUCATION REFORM EFFORT

The lessons (L) below are derived from my own interpretation of the physics-education reform movement and are therefore somewhat subjective and incomplete. They are meant to stimulate discussion rather than present any definitive final analysis.

### A. *Six Lessons On Interactive Engagement*

**L1. The use of Interactive Engagement (IE) strategies *can* increase the effectiveness of conceptually difficult courses well beyond that obtained with traditional methods.**
Education research in biology (Hake 1999a,b), chemistry (Herron & Nurrenbern 1999), and engineering (Felder et al. 2000a,b), although neither as extensive nor as systematic as that in physics (McDermott & Redish 1999, Redish 1999), is consistent with the latter in suggesting that in conceptually difficult areas, interactive engagement (IE) methods are more effective than traditional (T) passive-student methods in enhancing students' understanding. Furthermore, there is some preliminary evidence that learning in IE physics courses is substantially retained one to three years after the courses have ended (Chabay 1997, Francis et al. 1998, Bernhard 2000). I see no reason to doubt that enhanced understanding and retention would result from more use of interactive engagement methods in other science and even non-science areas, but substantive research on this issue is sorely needed – see L3 & L4.

**L2. The use of IE and/or high-tech methods, by themselves, does *not insure* superior student learning.**

As previously indicated, the data of Fig. 1 show that seven of the IE courses (717 students) achieved <g>'s close to those of the T courses. Five of those made extensive use of high-tech microcomuter-based labs (Thornton and Sokoloff 1990, 1998). Case histories of the seven low-<g> courses (Hake 1998b) suggest that implementation problems occurred.

Another example of the apparent failure of IE/high-tech methods has been described by Cummings et al. (1999). They considered a *standard* physics Studio Course at Rensselaer in which group work and computer use had been introduced as components of in-class instruction, the classrooms appeared to be interactive, and students seemed to be engaged in their own learning. Their measurement of <g>'s using the FCI and the Force Motion Concept Evaluation (Thornton & Sokoloff 1998) yielded values close to those characteristic of T courses (Hake 1998a,b,c). Cummings et al. suggest that the low <g> of the standard Rensselaer studio course may have been due to the fact that "the activities used in the studio classroom are predominately 'traditional' activities adapted to fit the studio environment and incorporate the use of computers." Thus the apparent "interactivity" was a product of traditional methods (supported by high technology), *not* published IE methods developed by physics-education researchers and based on the insights of cognitive scientists and/or outstanding classroom teachers, as for the survey courses. This explanation is consistent with the fact that Cummings et al. measured <g>'s in the 0.35 – 0.45 range for *revised* Rensselaer Studio courses using physics-education research methods: (a) Interactive Lecture Demonstrations (Thornton & Sokoloff (1998), and (b) Cooperative Group Problem Solving (Heller et al. 1992a,b) .

It should be emphasized that while high technology, by itself, is no panacea, it can be very advantageous when it promotes interactive engagement, as in, e.g.:

(a) computerized classroom communication systems [see, e.g., Dufresne et al. 1996**,** Mazur 1997, Abrahamson 1998, Burnstein & Lederman 2001, *Better Education* 2001];

(b) *properly implemented* microcomputer-based labs (Thornton and Sokoloff 1990);

(c) interactive computer animations for use *after* hands- and minds-on experiments and Socratic dialogue (Hake 2001a);

(d) computer implemented tutorials (Reif & Scott 1999);

(e) *Just-In-Time Teaching* (Novak et al. 1998, 1999; Gavrin 2001).

**L3. High-quality standardized tests of the cognitive and affective impact of courses are essential for gauging the relative effectiveness of non-traditional educational methods.**

As indicated in the introduction, so great is the inertia of the educational establishment (see L13) that three decades of physics-education research demonstrating the futility of the passive-student lecture in introductory courses were ignored until high-quality standardized tests that could easily be administered to thousands of students became available. These tests are yielding increasingly convincing evidence that interactive engagement methods enhance conceptual understanding and problem solving abilities far more than do traditional methods. Such tests may also indicate implementation problems in IE courses (Hake 1998b). As far as I know, disciplines other than physics, astronomy (Adams et al. 2000; Zeilik et al. 1997, 1998, 1999), and possibly economics (Saunders 1991, Kennedy & Siegfried 1997, Chizmar & Ostrosky 1998, Allgood and Walstad 1999) have yet to develop any such tests and therefore cannot effectively gauge either the need for or the efficacy of their reform efforts. In my opinion, *all disciplines should consider the construction of high-quality standardized tests of essential introductory course concepts.*

The lengthy and arduous process of constructing valid and reliable multiple choice tests has been discussed by Halloun & Hestenes (1985a), Hestenes et al. (1992), Beichner (1994), Aubrecht (1991), and McKeachie (1999). In my opinion such hard-won Diagnostic Tests that cover important parts of common introductory courses are national assets whose confidentiality should be as well protected as the MCAT (Medical College Admission Test). Otherwise the test questions may migrate to student files and thereby undermine education research that relies upon the validity of such tests. Suggestions for both administering Diagnostic Tests and reporting their results so as to preserve confidentiality and enhance assessment value have been given by Hake (2001b).

Regarding tests of *affective* impact:

> (a) administration of the *Maryland Physics EXpectations* (MPEX) survey to 1500 students in introductory calculus-based physics courses in six colleges and universities . . . . (showed). . . . "a large gap between the expectations of experts and novices and . . . . a tendency for student expectations to *deteriorate* rather than improve as a result of introductory calculus-based physics" (Redish et al. 1998). Here the term "expectations" is used to mean a combination of students' *epistemological* beliefs about learning and understanding physics and students' *expectations* about their physics course (Elby 1999). Elby (2001) has recently conducted classes so as to help students develop more sophisticated beliefs about knowledge and learning as measured by MPEX.

(b) The Arizona State University "Views About Sciences Survey" (VASS) (Halloun & Hestenes 1998, Halloun 1997) [available for physics, chemistry, biology and mathematics at < http://modeling.la.asu.edu/R&E/Research.html >] indicates that students have views about physics that (a) often diverge from physicists' views; (b) can be grouped into four distinct profiles: expert, high transitional, low transitional, and folk; (c) are similar in college and high school; and (d) *correlate significantly with normalized gain* g *on the FCI.* It may well be that students' attitudes and understanding of science and education are irreversibly imprinted in the early years [but see Elby (2001)]. If so, corrective measures await a badly needed shift of K-12 education away from rote memorization and drill (often encouraged by state-mandated standardized tests) to the enhancement of understanding and critical thinking (Hake 2000c,d; Mahajan & Hake 2000; Benezet 1935/36) – see L10.

**L4. Education Research and Development (R&D) by disciplinary experts (DE's), and of the same quality and nature as traditional science/engineering R&D, is needed to develop potentially effective educational methods within each discipline. But the DE's should take advantage of the insights of (a) DE's doing education R&D in other disciplines, (b) cognitive scientists, (c) faculty and graduates of education schools, and (d) classroom teachers.**

Redish (1999) has marshaled the arguments for the involvement of physicists in physics departments – not just faculty of education schools - in physics-education research. Similar arguments may apply to other disciplines. For physics, Redish gave these arguments: (a) physicists have good access to physics courses and students on which to test new curricula, (b) physicists and their departments directly benefit from physics education research, (c) education schools have limited funds for disciplinary education research, (d) understanding what's going on in physics classes requires deep rethinking of physics and the cognitive psychology of understanding physics. One might add that the researchers themselves must be excellent physics teachers with both content and "pedagogical content" knowledge (see L7) of a depth unlikely to be found among non-physicists.

The education of disciplinary experts in education research requires Ph.D. programs at least as rigorous as those for experts in traditional research. The programs should include, in addition to the standard disciplinary graduate courses, some exposure to: the history and philosophy of education, computer science, statistics, political science, social science, economics, engineering - see L11, and, most importantly, cognitive science (i.e., philosophy, psychology, artificial intelligence, linguistics, anthropology, and neuroscience). The breadth of knowledge required for effective education research is similar to that required in ecological research (Holling 1997). In the U.S. there are now about a dozen Ph.D. programs in physics education within physics departments and about half that number of interdisciplinary programs between physics and education or cognitive psychology (Physical Science Resource Center 2001, UMd-PERG 2001b.). In my opinion, *all scientific disciplines should consider offering Ph.D. programs in education research.*

But how can disciplinary education researchers, and for that matter, university faculty generally, take advantage of the insights of: disciplinary experts doing education R&D in other disciplines; cognitive scientists; faculty and graduates of education schools; and classroom teachers? (The current education-research schism between economics and physics is demonstrated in Sec. II-E.) In my opinion, even despite the rigid departmental separation of disciplines in most research universities, the web has the potential to dramatically enhance cooperation and interchange among these groups (Hake, 1999c, 2000e). Certainly the success of *Conservation Ecology* < http://www.consecol.org/Journal/ > testifies to the value of the web in promoting interdisciplinary understanding and effort. A starting point might be the construction of web guides for various disciplines similar to REDCUBE < http://www.physics.indiana.edu/~redcube > (Hake 1999b), which provides a point of entry into the vast literature and web resources relevant to *REsearch, Development, and Change in Undergraduate Biology Education*. The 9/8/99 version contains 47 biology-educator profiles; 446 references (including 124 relevant to general science-education reform); and 490 hot-linked URL's on (a) Biology Associations, (b) Biology Teachers' Web Sites, (c) Scientific Societies and Projects (not confined to Biology), (d) Higher Education, (e) Cognitive Science and Psychology, (f) U.S. Government, and (g) Searches and Directories.

**L5. The development of effective educational methods within each discipline requires a redesign process of continuous long-term classroom use, feedback, assessment, research analysis, and revision.**

Wilson and Daviss (1994) suggest that the "redesign process," used so successfully to advance technology in aviation, railroads, automobiles, and computers can be adapted to K-12 education reform through "System Redesign Schools." Redesign processes in the reform of introductory undergraduate physics education have been undertaken and described by McDermott (1991) and by Hake (1998a). In my opinion "redesign" at both the K-12 and undergraduate levels can be greatly assisted by the promising *Scholarship of Teaching & Learning* movement (Carnegie Academy 2000) inspired by Boyer (1990) and the Boyer Commission (1998).

**L6. Although non-traditional interactive-engagement methods appear to be much more effective than traditional methods, there is need for more research to develop better strategies for the enhancement of student learning.**

On a test as elemental as the FCI it would seem that reasonably effective courses should yield $<g>$'s above 0.8, but thus far none much above 0.7 have, to my knowledge, been reported. This and the poor showing on the pre/post MPEX test of student understanding of the nature of science and education (Redish et al. 1998) indicates that more work needs to be done to improve IE methods. It would seem that understanding of science might be improved by:

- (a) students' apprenticeship research experiences (Collins et al. 1989, Brown et al. 1989);
- (b) epistemolgically oriented teachers, materials, and class activities (Elby 2001); and
- (c) enrollment in courses featuring interactive engagement among students and disciplinary experts from different fields, *all in the same classroom at the same time* (Benbasat & Gass 2001).

In my opinion, more support should be given by universities, foundations, and governments to the development of a *science of education* spearheaded by *disciplinary* education researchers working in concert with cognitive scientists and education specialists. In the words of cognitive psychologists Anderson et al. (1998): "The time has come to abandon philosophies of education and turn to a *science of education* . . . . . . If progress is to be made to a more scientific approach, traditional philosophies . . . .(such as radical constructivism) . . . . will be found to be like the doctrines of folk medicine. They contain some elements of truth and some elements of misinformation . . . . . . *Only when a science of education develops that sorts truth from fancy - as it is beginning to develop now will dramatic improvements in educational practice be seen.*" (My italics.)

J.J. Duderstadt (2001), president emeritus of the University of Michigan and chair of the National Academies *Committee on Science, Education, and Public Pol*icy (COSEPUP) < http://www4.nationalacademies.org/pd/cosepup.nsf >, cogently argues that "the development of human capital is becoming a dominant national priority in the age of knowledge, comparable in importance to military security and health care. Yet our federal investment in the knowledge base necessary to address this need in miniscule. In FY01. . . (Fiscal Year 2001). . . the nation will invest over $247 billion in R&D . . . . *How much will the federal government invest in research directed toward learning, education, and schools? Less than $300 million* . . . . most industries spend between 3% to 10% per year of revenues for R&D activities. By this measure, the education sector of our economy (including K-12, higher education, and workforce training), which amounts to $665 billion, should be investing $20 billion or greater each year in R&D, roughly the same order of magnitude as the health care sector. . . . .(Focusing on). . . what many term the '*science of education*,' meaning *research* *that would be classified by scientists as guided by the scientific method and subject to rigorous review by the scientific community. . .* (See Sec. II-G). . . an interesting model for the conduct of research on education and learning is provided by the DOD's Defense Advanced Research Programs Agency (DARPA). Through a process using visionary program managers to channel significant, flexible, and long-term funding to the very best researchers for both basic and applied research undergirding key defense technologies, DARPA has been able to capture contributions of the very best of the nation's scientists and engineers in highly innovative projects. . . . . Perhaps we need an *Education Advanced Research Programs Agency* . . .(EARPA). . . . to focus the capabilities of the American research enterprise on what many believe to be our nation's most compelling priority, the quality of education for a knowledge-driven society. . . .  If the past 50 years of science policy can be characterized as a transition in national priorities 'from guns to pills,' let me suggest that *the next 50 years will see the transition 'from pills to brains.'  It is time that we realized that our nation's intellectual capital, the education of our people, the support of their ideas, their creativity, and their innovation, will become the dominant priority of a knowledge-driven nation . . .* " (My *italics*.)

However, it should be emphasized that the development of better strategies for the enhancement of student learning will not improve the educational system unless (a) university and K-12 teachers (see L10) are educated to effectively implement those strategies, and (b) research universities start to think of education in terms of *student learning* rather than the *delivery of instruction* (see L12h).  In Duderstadt's (2001) words: "Beyond new mechanisms to stimulate and support research in the science of education, *we also need to develop more effective mechanisms to transfer what we have learned into schools, colleges, and universities.* For example, the progress made in cognitive psychology and neuroscience during the past decade in the understanding of learning is considerable. Yet almost none of this research has impacted our schools. As one of my colleagues one said, "*If doctors used research like teachers do, they would still be treating patients with leeches*." (My *italics*.)

**B.** *Eight Lessons On Implementation*

**L7. Teachers who possess *both* content knowledge and "pedagogical content knowledge" are more apt to deliver effective instruction.**

"Pedagogical content knowledge" is evidently a term due to Shulman (1986, 1987), but its importance has long been well known to effective classroom teachers. The difference between content knowledge and "pedagogical content knowledge," can be illustrated by consideration of the HH-type question given in the Introduction. *Content knowledge* informs the teacher that, according to Newton's First Law, while the brick is moving vertically upward at a constant speed in the inertial reference frame of the lab, the magnitude of the force on the brick by the student's hand is constant in time and of magnitude W, so that the *net* force on the brick is zero. On the other hand, *pedagogical content knowledge* would inform the teacher that students may think that e.g.: (a) since a net force is required to produce motion, the force on the brick by the student's hand is constant in time and greater than W; or (b) since the weight of the brick diminishes as it moves upward away from the Earth, the force on the brick by the student's hand decreases in time but is always greater than W; or (c) no force is exerted on the brick by the student's hand because as the students hand moves up the brick must simply move up to stay out of the hand's way. In addition, pedagogical content knowledge provides a hard-won toolkit of strategies (see, e.g., the list of "Popular IE Methods" in II-C above) for guiding the student away from these misconceptions and towards the Newtonian interpretation. Unfortunately, such knowledge may take many years to acquire (Wells et al. 1995).

**L8. College and university faculty tend to overestimate the effectiveness of their own instructional efforts and thus tend to see little need for educational reform.**

As examples of this tendency see Geilker (1997) [countered by Hilborn (1998)]; Griffiths (1997) [countered by Hestenes (1998)]; Goldman (1998); Mottman (1999a,b) [countered by Kolitch (1999), Steinberg (1999), and Hilborn (1999)]; and Carr (2000).

**L9. Such complacency can sometimes be countered by the administration of high-quality standardized tests of understanding and by "video snooping."**

   a. Harvard's Eric Mazur (1997) was very satisfied with his introductory-course teaching - he received very positive student evaluations and his students did reasonably well on "difficult" exam problems. Thus it came as a shock when his students fared hardly better on the "simple" FCI than on their "difficult" midterm exam. As a result, Mazur developed and implemented his interactive- engagement *Peer Instruction* method as a replacement for his previous traditional passive-student lectures. This change resulted in much higher $<g>$'s on the FCI as shown by comparison of the red and green triangular points with average pretest scores in the vicinity of 70% in Fig. 1.

   b. Like Mazur, most Harvard faculty members are proud of their undergraduate science courses. However, the videotape *Private Universe* (Schneps & Sadler 1985) shows Harvard graduating seniors being asked "What causes the seasons?" Most of them confidently explain that the seasons are caused by yearly changes in the distance between the Sun and the Earth! Similarly most MIT faculty regard their courses as very effective preparation for the difficult engineering problems that will confront their elite graduates in professional life. However the videotape *Simple Minds* (Shapiro et al. 1997) shows MIT graduating seniors having great trouble getting a flashlight bulb to light, given one bulb, one battery, and one piece of wire.

**L10. A major problem for undergraduate education in the United States is the inadequate preparation of incoming students, in part due to the inadequate university education of K-12 teachers.**

   According to the National Research Council (1999), the Third International Mathematics and Sciences Survey (TIMSS) indicates that: "U.S. students' worst showing was in population 3 . . . . (final year of secondary School. . . . corresponding to U.S. high school seniors). . . . In the assessment of general mathematics and science knowledge, U.S. high school seniors scored near the bottom of the participating nations. In the assessments of advanced mathematics and physics given to a subset of students who had studied those topics, no nations had significantly lower mean scores than the United States. The TIMSS results indicate that a considerably smaller percentage of U.S. students meet high performance standards than do students in other countries." Consistent with the foregoing, I have observed (Hake 2000d) that FCI pretest averages for students entering the introductory physics course at Indiana University are quite low (30% - 45%) and about the same regardless of whether or not the students are graduates of high-school physics classes.

But it's not just a matter of physics floundering. According to Epstein (1997-98): "While it is now well known that large numbers of students arrive at college with large educational and cognitive deficits many faculty and administrative colleagues are not aware that many students lost all sense of meaning or understanding in elementary school……In large numbers our students …… [at Bloomfield College (New Jersey) and Lehman (CUNY)] ….. cannot order a set of fractions and decimals and cannot place them on a number line. Many do not comprehend division by a fraction and have no concrete comprehension of the process of division itself. Reading rulers where there are other than 10 subdivisions, basic operational meaning of area and volume, are pervasive difficulties. Most cannot deal with proportional reasoning nor any sort of problem that has to be translated from English. Our diagnostic test, which has now been given at more than a dozen institutions shows that there are such students everywhere" . . . . . .[even Wellesley (Epstein 1999)].

Kati Haycock (1999), director of the American Association of Higher Education's (AAHE's) Education Trust  < http://www.edtrust.org/ > hits the nail on the head: "Higher education…. (unlike Governors and CEO's) ….. has been left out of the loop and off the hook …. (in the effort to improve America's public schools since release of *A Nation at Risk* in 1983)…. Present neither at the policy tables where improvement strategies are formulated nor on the ground where they are being put into place, most college and university leaders remain blithely ignorant of the roles their institutions play in helping K-12 schools get better - and the roles they currently play in maintaining the status quo …. *How are we going to get our students to meet high standards if higher education continues to produce teachers who don't even meet those same standards*?  How are we going to get our high school students to work hard to meet new, higher standards if most colleges and universities will continue to admit them regardless of whether or not they even crack a book in high school?"  (My *italics*.)

According to the NSF Advisory Committee (1996): "Many faculty in SME&T. . . . (Science, Math, Engineering, and Technology) . . . .  at the post-secondary level continue to blame the schools for sending underprepared students to them.  But, increasingly, the higher education community has come to recognize the fact that teachers and principals in the K-12 system are all people who have been educated at the undergraduate level, mostly in situations in which *SME&T programs have not taken seriously enough their vital part of the responsibility for the quality of America's teachers*."  (My *italics*.) See also NSF Advisory Committee (1998).

Fortunately, despite the general failure of *pre-service* teacher education, several programs have been established over the past few years to enhance the pedagogical skills and content knowledge of *in-service* physics teachers. For a hot-linked list of 25 such programs see Hake (2000d).

The Glenn Commission (2000) proposals may be a step in the right direction. The commission requests *5 billion dollars* in the first year to initiate (my *italics*):
- a. establishment of an *ongoing system to improve the quality of mathematics and science teaching in grades K–12*,
- b. significant increase in the number of mathematics and science teachers with *improved quality of their preparation*,
- c. improvement of the working environment and so as to *make the teaching profession more attractive for K–12 mathematics and science teachers*.

More recently, the U.S. National Security – Hart-Rudman Commission (2001) has warned that " . . . *the U.S. need for the highest quality human capital in science, mathematics, and engineering is not being met.* . . (partially because) . . . . the American . . . (K-12) . . . education system is not performing as well as it should," and recommends a *National Security Science and Technology Education Act* to fund a comprehensive program to produce the needed numbers of science and engineering professionals as well as qualified teachers in science and math."

**L11. Interdisciplinary cooperation of instructors, departments, institutions, and professional organizations is required for synthesis, integration, and change in the entire chaotic educational system.**
>Although more research to develop better strategies for the enhancement of student learning (L6) is required, that by itself will not reform the entire chaotic educational system, as has been emphasized by Tobias (1992a,b; 2000), Sarason (1990, 1996), Hilborn (1997), and Wilson & Daviss (1994). In my opinion, an *engineering* approach to the improvement of education (Felder 2000a,b) seems to be required. Bordogna (1997) conveys the essence of engineering as "integrating all knowledge for some purpose. . . . The engineer must be able to work across many different disciplines and fields - and make the connections that will lead to deeper insights, more creative solutions, and getting things done. In a poetic sense, paraphrasing the words of Italo Calvino (1988), *the engineer must be adept at correlating exactitude with chaos to bring visions into focus.*" (My *italics*).  It would appear that "engineering" as seen by Bordogna is similar to "integrative science" as seen by Holling (1998).

**L12. Various institutional and political factors, including the culture of research universities, slow educational reform.** Those listed below pertain to the United States, but similar barriers may exist in other countries.
>Among the institutional and political factors listed by Tobias (2000) as thwarting educational reform are (those most associated with the culture of research universities are indicated in *italics*):
>>a. Advanced Placement (AP) courses serve as a filter rather than a pump.
>>b. *In-class and standardized tests (MCAT, SAT, GRE) drive the curriculum in a traditional direction*.
>>c. *Effectiveness of teaching has little effect on promotion/tenure decisions or on national departmental rankings*.
>>d. High-school science courses are not required for college admission; many colleges require little or no science for graduation.
>>e. *Clients for the sciences aren't cultivated among those who do not wish to obtain PhD.*'s.
>>f. *Class sizes are too large*.

To Tobias's list I would add:

    g. The failure of the K-12 system to incorporate physics – the most basic of the sciences and essential for any proper understanding of biology and chemistry – into *all* grades for *all* students (Ford 1989, Swartz 1993, Hammer 1999, Neuschatz 1999, Lederman 1999, Livanis 2000). In the words of physics Nobelist Leon Lederman: "We have observed that 99 percent of our high schools teach biology in 9th (or 10th) grade, chemistry in 10th or 11th grade, and, for survivors, physics in 11th or 12th grade. This is alphabetically correct, but by any logical scientific or pedagogical criteria, the wrong order. . . . This reform . . . .("physics first"). . . . concentrates on installing a coherent, integrated science curriculum, which matches the standards of what high school graduates should understand and be able to do . . . . And wouldn't it be a natural next step to invite the history teachers, the teachers of arts and literature, to help develop those connections of the fields of learning that the biologist E.O. Wilson (1998) calls 'consilience'?" Arons (1959) took an early step in this direction at Amherst, but his attempts to bridge the "two-culture gap" were abandoned soon after his departure. For some other attempts to link science and liberal-arts education see e.g., Tobias & Hake (1988), Tobias & Abel (1990), and the AAAS (1990) report on the "liberal art of science."

    h. *The failure of research universities to:*

      *(1) Discharge their obligation to adequately educate prospective K-12 teachers* (Hake 2000c,d) – see L10.

     (2) *Think of education in terms of <u>student learning</u> rather than the <u>delivery of instruction</u>* (Barr & Tagg 1995; Duderstadt 2000, 2001). An emphasis on the *learning* paradigm may be encouraged by:

        (a) the previously mentioned *Scholarship of Teaching & Learning* movement (Carnegie Academy 2000) inspired by Boyer (1990) and the Boyer Commission (1998);

        (b) the National Academy for Academic Leadership < http://www.thenationalacademy.org/ >, which strives to "educate academic decision makers to be leaders for sustained, integrated institutional change that significantly improves student learning";

        (c) threats from accrediting agencies such as ABET (Accreditation Board for Engineering and Technology < http://www.abet.org/ >) with its emphasis on *accountability for actual student learning* (Van Heuvelen & Andre 2000; Heller 2000; Hake 2000c,d); and

        (d) competition for transmission-mode lecture services from distance-education conglomerates (Marchese 1998, Duderstadt 2000).

(3) *Effectively consider crucial multidisciplinary societal problems such as education*.

In the words of Karl Pister (1996), former Chancellor of UC - Santa Cruz: ". . . we need to encourage innovative ways of looking at problems, moving away from the increasing specialization of academia to develop new interdisciplinary fields that can address complex real-world problems from new perspectives."

i. The failure of society to pay good K-12 teachers what they are worth. Physicist Don Langenberg (1999), chancellor of the University System of Maryland and president of the National Association of System Heads < http://www.nashonline.org/ >, suggests that "on average, *teacher's salaries ought to be about 50% higher than they are now*. Some teachers, including the very best, those who teach in shortage fields (e.g., math and science) and those who teach in the most challenging environments (e.g., inner cities) ought to have salaries about twice the current norm . . . . Simple arithmetic applied to publicly available data shows that the increased cost would be only 0.6% of the GDP. . . .(i.e., about 600 billion dollars over 10 years). . . . about one twentieth of what we pay for health care. *I'd assert that if we can't bring ourselves to pony up that amount, we will pay far more dearly in the long run*." (My *italics*.) A similar proposal with a similar cost estimate (about 450 billion dollars over 10 years) has been made independently by physicist Ken Heller (2001). More restrictively, the United States Commission on National Security/21st Century - Hart-Rudman Commission (2001b) estimates a cost of 64 billion dollars over 10 years to raise the salaries of all public secondary school science and math teachers such that average yearly starting salaries would be raised from the current $25,000 to $50,000.

**L13. The monumental inertia of the educational system may thwart long-term national reform.**

The glacial inertia of the nearly immovable U.S. educational system is not well understood. A recent issue of *Daedalus* (1998) contains essays by researchers in education and by historians of more rapidly developing institutions such as power systems, communications, health care, and agriculture. The issue was intended to help answer a challenge posed by physics Nobelist Kenneth Wilson: "If other major American 'systems' have so effectively demonstrated the ability to change, why has the education 'system' been so singularly resistant to change? What might the lessons learned from other systems' efforts to adapt and evolve have to teach us about bringing about change - successful change – in America's schools?" As far as I know, no definitive answer has yet been forthcoming.

Clifford Swartz (1999), former editor of *The Physics Teacher* and long-time acerbic critic of physics-education research, wrote: " There is a variety of evidence, and claims of evidence, that each of the latest fads . . .(constructivism, 'group' and 'peer' instruction, 'interaction') . . . produces superior learning and happier students. In particular, students who interact with apparatus or lecture do better on the *Force Concept Inventory* exam (Hestenes et al. 1992). The evidence of Richard Hake's (1998a) metastatistical study is so dramatic that the only surprising result is that many schools and colleges are still teaching in old-fashioned ways. Perhaps the interaction technique reduces coverage of topics, or perhaps the method requires new teaching skills that teachers find awkward. *At any rate the new methodology is not sweeping the nation*." (My italics.)

New educational methodologies *have* from time to time swept the nation (e.g., "the new math," PSSC (Physical Science Study Committee) physics, the Keller Plan (Personalized System of Instruction) but then faded from sight. History (Holton 1986; Arons 1993, 1997, 1998); Sarason 1990, 1996; Cuban 1999) suggests that the present educational reform effort may, like its predecessors, have little lasting impact. This would be most unfortunate, considering the current imperative to:
   a. educate more effective science majors and science-trained professionals,
   b. raise the appallingly low level of science literacy among the general population,
   c. *solve the monumental science-intensive problems (economic, social, political, and environmental) that beset us.*

**L14. "Education is not rocket science, it's much harder."**
George Nelson, astronaut and astrophysicist, as quoted by Redish (1999).

My own belief, conditioned by 40 years of research in superconductivity and magnetism, 28 years in physics teaching, and 16 years in education research, is that *effective* education (both physics teaching and education research) is harder than solid-state physics.  The latter is, of course, several orders of magnitude harder than rocket science. Nuclear physicist Joe Redish (1999) writes: "The principles of our first draft of a community map for physics education are different in character from the laws we would write down for a community map of the physical world.  They are much less like mathematical theorems and much more like heuristics.  This is not a surprise, since the phenomena we are discussing are more complex and at a much earlier stage of development."  Since education is a complex, early-stage, dynamic, non-linear, scientific/sociopolitical, high-stakes system, it might benefit from the expertise of conservation ecologists who are well used to dealing with such challenging systems. (Holling 1999).

---

## RESPONSES TO THIS ARTICLE

Responses to this article are invited.  If accepted for publication, your responses will be hyperlinked to the article.  To submit a comment, follow [this link](#).  To read comments already accepted, follow [this link](#).

---

**Acknowledgements:**

*I should like to dedicate this paper to the late Arnold Arons, farsighted pioneer of U.S. physics education research and the major source of wisdom, educational inspiration, and encouragement to me (Hake 1991) and many others over the decades.  I thank David Hestenes for insights and assistance, Werner Wittmann for sage comments on statistics, Bill Becker for his stimulating econometric perspectives, and a discerning referee for excellent suggestions that improved this article. I am also indebted to Lee Gass for suggesting that I write this review, and for his very valuable comments on the manuscript. Finally, I thank the National Science Foundation for funding through NSF Grant DUE/MDR-9253965.*

**Burnstein, R.A. & L.M. Lederman.** 2001. Using wireless keypads in lecture classes. *Phys. Teach.* **39**(1):8-11.

**Calvino, I.** 1988. *Six memos for the next millennium*. Harvard University Press.

**Carnegie Academy.** 2000. Scholarship of teaching and learning. [online] URL: < http://www.carnegiefoundation.org/CASTL/index.htm >.

**Carr, J.J.** 2000. The physics tutorial: some cautionary remarks. *Am. J. Phys.* **68**(11):977-978.

**Carver. R.P.** 1993. The case against statistical significance testing, revisited. *Journal of Experimental Education* **61**(4): 287-292.

**Chabay, R.W.** 1997. Qualitative understanding and retention. *AAPT Announcer* **27**(2):96.

**Chizmar, J.F. & A.L. Ostrosky.** 1998. The one-minute paper: Some empirical findings. *Journal of Economic Education*. Winter **29**(1):3-10. [online] URL: < http://www.indiana.edu/~econed/issues/v29_1/1.htm >.

**Clement, J.M.** 2001. The correlation between scientific thinking skill and gain in physics conceptual understanding. *AAPT Announcer* **31**(2):82.

**Cohen, J.** 1988. *Statistical power analysis for the behavioral sciences*. Lawrence Erlbaum, 2nd ed.

**Cohen, J.** 1994. The earth is round (*p* < .05). *American Psychologist* **49**:997-1003.

**Collins, A., J.S. Brown, and S. Newman.** 1989. Cognitive apprenticeship: teaching students the craft of reading, writing, and mathematics. In L.B. Resnick, ed*., Knowing, learning, an instruction: Essays in honor of Robert Glaser*, pp. 453-494. Lawrence Erlbaum.

**Cook, T.D., & D.T. Campbell.** 1979. *Quasi-experimentation: design & analysis issues for field settings.* Houghton Mifflin.

**Cronbach, L.J. & L. Furby**. 1970. How should we measure "change" – or should we? *Psychological Bulletin* **74**:68-80.

**Crouch, C.H. & E. Mazur,** 2001. Peer instruction: Ten years of experience and results. *Am. J. Phys.,* in press.
44

**Address of Correspondent:**
Richard R. Hake
Emeritus Professor of Physics, Indiana University
24245 Hatteras Street, Woodland Hills, CA 91367 USA
Phone: 818-992-0632
<rrhake@earthlink.net>
< http://www.physics.indiana.edu/~hake/ >